\documentclass[prd,reprint,nofootinbib,notitlepage,aps,tightenlines,
preprintnumbers,amsmath,amssymb,showpacs,superscriptaddress]{revtex4-1}

\usepackage[hyperindex,breaklinks]{hyperref}
\usepackage{graphicx,epstopdf,amsmath,amsfonts,amssymb,color}

\def\be{\begin{equation}}
\def\ee{\end{equation}}
\def\ba{\begin{eqnarray}}
\def\ea{\end{eqnarray}}
\def\ge{\mathrel{\raise.3ex\hbox{$>$\kern-.75em\lower1ex\hbox{$\sim$}}}}
\def\la{\mathrel{\raise.3ex\hbox{$<$\kern-.75em\lower1ex\hbox{$\sim$}}}}

\def\simgt{\mathrel{\raise.3ex\hbox{$>$\kern-.75em\lower1ex\hbox{$\sim$}}}}
\def\simlt{\mathrel{\raise.3ex\hbox{$<$\kern-.75em\lower1ex\hbox{$\sim$}}}}

\newcommand{\bi}[1]{\bibitem{#1}}
\newcommand{\fr}[2]{\frac{#1}{#2}}

\newcommand{\nc}{\newcommand}

\nc{\gone}{\bar g_{\pi NN}^{(1)}}
\nc{\gzero}{\bar g_{\pi NN}^{(0)}}
\nc{\al}{\alpha}
\nc{\ga}{\gamma}
\nc{\de}{\delta}
\nc{\ep}{\epsilon}
\nc{\ze}{\zeta}
\nc{\et}{\eta}
\nc{\ka}{\kappa}
\nc{\rh}{\rho}
\nc{\si}{\sigma}
\nc{\ta}{\tau}
\nc{\up}{\upsilon}
\nc{\ph}{\phi}
\nc{\ch}{\chi}
\nc{\ps}{\psi}
\nc{\om}{\omega}
\nc{\Ga}{\Gamma}
\nc{\De}{\Delta}
\nc{\La}{\Lambda}
\nc{\Si}{\Sigma}
\nc{\Up}{\Upsilon}
\nc{\Ph}{\Phi}
\nc{\Ps}{\Psi}
\nc{\Om}{\Omega}
\nc{\ptl}{\partial}
\nc{\del}{\nabla}
\nc{\ov}{\overline}
\nc{\newcaption}[1]{\centerline{\parbox{15cm}{\caption{#1}}}}
\nc{\us}{U(1)$_S$}
\nc{\Rg}{$R_{\gamma\gamma}$}

\def\beq{\begin{equation}}
\def\eeq{\end{equation}}
\def\bmat{\begin{displaymath}}
\def\emat{\end{displaymath}}
\def\bear{\begin{eqnarray}}
\def\eear{\end{eqnarray}}
\def\ba{\begin{eqnarray}}
\def\ea{\end{eqnarray}}
\def\bery{\begin{array}}
\def\ery{\end{array}}
\def\bit{\begin{itemize}}
\def\eit{\end{itemize}}
\def\ben{\begin{enumerate}}
\def\een{\end{enumerate}}
\def\btab{\begin{tabular}}
\def\etab{\end{tabular}}
\def\btbl{\begin{table}}
\def\etbl{\end{table}}
\def\bfig{\begin{figure}[htb]}
\def\efig{\end{figure}}
\def\bpic{\begin{picture}}
\def\epic{\end{picture}}

\def\ga{\mathrel{\raise.3ex\hbox{$>$\kern-.75em\lower1ex\hbox{$\sim$}}}}
\def\la{\mathrel{\raise.3ex\hbox{$<$\kern-.75em\lower1ex\hbox{$\sim$}}}}
\def\gappeq{\mathrel{\rlap {\raise.5ex\hbox{$>$}}
{\lower.5ex\hbox{$\sim$}}}}
\def\lappeq{\mathrel{\rlap{\raise.5ex\hbox{$<$}}
{\lower.5ex\hbox{$\sim$}}}}

\def\gyr{{\rm \, G\kern-0.125em yr}}
\def\mev{{\rm \, Me\kern-0.125em V}}
\def\gev{{\rm \, Ge\kern-0.125em V}}
\def\tev{{\rm \, Te\kern-0.125em V}}

\def\lsim{\mathrel{\rlap{\lower4pt\hbox{\hskip1pt$\sim$}}
    \raise1pt\hbox{$<$}}}                
\def\gsim{\mathrel{\rlap{\lower4pt\hbox{\hskip1pt$\sim$}}
    \raise1pt\hbox{$>$}}}                

\begin{document}
 
\title{Modified Higgs branching ratios versus $CP$ and lepton flavor violation}

\author{David McKeen}
\affiliation{Department of Physics and Astronomy, University of Victoria, 
Victoria, BC V8P 5C2, Canada}

\author{Maxim Pospelov}
\affiliation{Department of Physics and Astronomy, University of Victoria, 
Victoria, BC V8P 5C2, Canada}
\affiliation{Perimeter Institute for Theoretical Physics, Waterloo, ON N2J 2W9, 
Canada}

\author{Adam Ritz}
\affiliation{Department of Physics and Astronomy, University of Victoria, 
Victoria, BC V8P 5C2, Canada}

\date{August 2012}

\begin{abstract}
New physics thresholds which can modify the diphoton and dilepton Higgs branching ratios significantly may also provide new sources of $CP$ and lepton flavor violation. We find that limits on electric dipole moments impose strong constraints on any $CP$-odd contributions to Higgs diphoton decays unless there are degeneracies in the Higgs sector that enhance $CP$-violating mixing. We exemplify this point in the language of effective operators and in simple UV-complete models with vector-like fermions. In contrast, we find that electric dipole moments and lepton flavor violating observables provide less stringent constraints on new thresholds contributing to Higgs dilepton decays.

\end{abstract}
\maketitle

\section{Introduction}
\label{sec:intro}
The recent discovery of a Higgs-like resonance at the LHC \cite{Higgs}, with a mass of approximately $125$~GeV consistent
with electroweak precision observables,  has solidified the impressive verification of the Standard Model (SM)
at the electroweak scale. At the present time, the couplings of this resonance agree  on average rather well with those of 
the SM Higgs boson. 

The lack of hints for New Physics (NP) in other channels has focused attention on the detailed properties
of the Higgs-like resonance, and deviations from the SM in its decays to various final states. Indeed, while
the LHC now strongly constrains NP that can be produced either resonantly or in pairs from proton constituents
with well-identifiable final states---e.g., $Z^\prime$ bosons decaying to leptons, or squark/gluino decays to jets, leptons and
missing energy---NP produced via electroweak interactions or 
other weakly coupled hidden sectors is far less constrained. 
The latter possibilities are now coming under additional scrutiny
as possible explanations for small $2\sigma$ deviations from the SM  
in certain Higgs production/decay channels \cite{Higgs}, specifically, the apparent enhancement in the diphoton branching Br$(h\rightarrow \gamma\gamma)$~\cite{gammagrave} and a
possible suppression of decays to dileptons in Br$(h\rightarrow \tau\tau)$. Although these
deviations are small and may well dissipate with more data, they motivate the exploration of  viable models of NP that
could provide an explanation. The recent literature has focused on Br$(h\rightarrow \gamma\gamma)$ and noted that relatively light
(typically sub 300~GeV) electromagnetically charged fields that are vector-like (VL), i.e., with 
a contribution to their mass which does not come from electroweak symmetry breaking, can lead
to the required enhancement while still being accessible with sufficient statistics at the LHC \cite{recentVL,VL_EWP,bmp}.

Exploration of Higgs interactions in this way will be an important probe of NP in coming years, and thus it
is important to clarify the full range of interactions that allow for measurable corrections to the Higgs branching rates,
and the interplay with other precision data, particularly in the Yukawa sector.
In this paper, we ask whether new VL thresholds contributing to sizable deviations from SM Higgs branching can
also provide new sources of $CP$ and flavor violation \cite{prehistory,oldstuff}.  In Sec.~\ref{sec:edm}, building on \cite{oldstuff} 
we focus on the $CP$-odd operator $h F_{\mu\nu}\tilde F_{\mu\nu}$, and elucidate the connection between the $CP$-violating
Higgs decay amplitude and the impressive constraints on the electric dipole moments (EDMs) of elementary particles
\cite{YbF,Tl,Hgnew,n}. We find that the inferred bound on the EDM of the electron \cite{YbF,Tl} does not allow for
significant $CP$-odd contributions to the Higgs diphoton decay at the level of this dimension-five operator. We then 
consider two UV completions involving VL fermions and/or singlets, and identify a special case where the Higgs is nearly degenerate
with a singlet scalar that allows for large $CP$-odd contributions to the diphoton decay that can escape EDM bounds. In Sec.~\ref{sec:flavor}, 
we turn our attention to the $(H^\dagger H) \bar{L}_L^i H e^j_R$ operators contributing to dilepton decays, and consider the benchmark  
sensitivity from lepton flavor-violating (LFV)  observables and EDMs. Section~\ref{sec:conclusions} contains some concluding remarks.

\section{EDMs vs  diphoton decays}
\label{sec:edm}
Consider new physics charged under SU(2)$\times$U(1) only, so that the 
leading dimension-six operators which correct the diphoton branching ratio of the Higgs are
\begin{align}
 \De {\cal L} &= \frac{g_1^2}{e^2\La^2} H^\dagger H \left( a_h B_{\mu\nu} B^{\mu\nu} +\tilde{a}_h B_{\mu\nu} \tilde{B}^{\mu\nu}\right)
\nonumber
\\
&~~~~ + \frac{g_2^2}{e^2\tilde{\La}^2} H^\dagger H \left( b_h W_{\mu\nu} W^{\mu\nu} + \tilde{b}_h W_{\mu\nu}\tilde{W}^{\mu\nu}\right) \\
  &\rightarrow \frac{c_h v}{\La^2} h F_{\mu\nu}F^{\mu\nu} + \frac{\tilde{c}_h v}{\tilde{\La}^2} h F_{\mu\nu} \tilde{F}^{\mu\nu} + \cdots \label{Leff}
\end{align}
Here $c_h=a_h+b_h$, $\tilde{c}_h=\tilde{a}_h + \tilde{b}_h$, $v=246$~GeV and we have only retained 
the $h\gamma\gamma$ operators, disregarding couplings to $Z$ and $W$. Since we focus on
corrections that are sizable for loop-induced couplings to the photon, the associated corrections to the tree-level $hZZ$ and $hWW$ couplings
can be consistently ignored.\footnote{For recent studies of the $CP$ properties of the $hZZ$ and $hWW$ couplings, see, e.g.,~\cite{hVV,new_param} and references therein.} 
For thresholds in the TeV range or above, measurement of the Higgs decay rate itself probably provides the best sensitivity to $\La$. However,
EDMs can provide sensitivity to the $CP$-odd threshold $\tilde{\La}$.

The ensuing correction to the SM $h\to \gamma\gamma$ width, 
\be
\Ga_{\gamma\gamma}^{\rm SM} =\frac{m_h^3}{4\pi} \left( \fr{\alpha}{4\pi}\right)^2\left| \fr{ A_{\rm SM}}{2v} \right |^2 \simeq 9.1~{\rm keV},
\ee
takes the form
\be
R_{\gamma\gamma} = \frac{\Gamma_{\gamma\gamma}}{\Ga_{\gamma\gamma}^{\rm SM}} \simeq \left|1-{c}_h\frac{v^2}{\La^2}\frac{8 \pi }{\al A_{\rm SM}}\right|^2 + 
\left|\tilde{c}_h\frac{v^2}{\tilde\La^2}\frac{8 \pi }{\al A_{\rm SM}}\right|^2,
\ee
where $A_{\rm SM}(m_h=125~{\rm GeV}) \simeq A_W +A_t \simeq -6.5$ is proportional to the SM amplitude \cite{Djouadi}.  The deviations in the width are of ${\cal O}(1)$ for $\La/\sqrt{c_h}\sim 5$~TeV. Note that
since the $CP$-odd operator does not interfere with the SM amplitude, 
the corresponding correction to the diphoton branching ratio is necessarily positive and scales as ${\cal O}(1/\tilde{\La}^4)$.

\subsection{EDM limit on contact operators}
\label{sec:contact_op_lim}

Current experiments \cite{YbF,Tl,Hgnew,n}  already probe the EDMs of elementary particles at a level roughly commensurate 
with two-loop electroweak diagrams  \cite{PR}, with the chirality of light particles protected by factors of $m_{e(q)}/v$. Thus it is
useful to introduce the auxiliary quantity $d^{(2l)}_f$ that quantifies this  two-loop benchmark EDM scale, 
\be
d^{(2l)}_{f} \equiv \frac{|e| \alpha m_f}{16\pi^3 v^2} \;\;\; \Longrightarrow \;\;\; d^{(2l)}_{e} \simeq 2.5\times 10^{-27}~e\cdot {\rm cm}.
\ee
One observes that  $d^{(2l)}_{e}$ has already been surpassed by the current electron EDM limits \cite{YbF,Tl}, with the mercury \cite{Hgnew}
and neutron \cite{n} EDMs not lagging far behind for $d_q^{(2l)}$ \cite{PR}. 

\begin{figure*}
\includegraphics[scale=0.7]{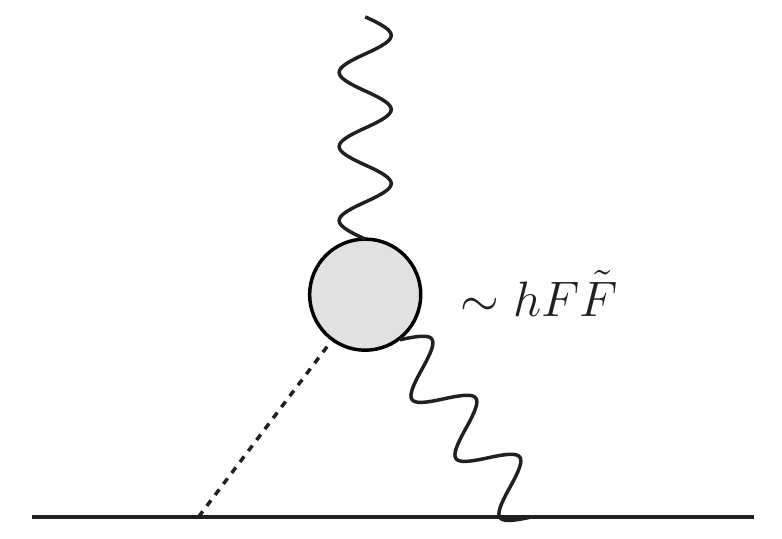}\hspace*{0.5cm}\includegraphics[scale=0.7]{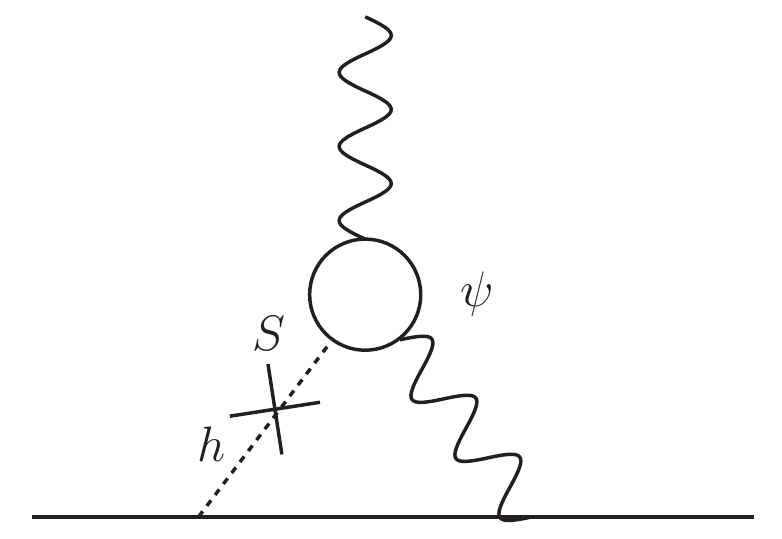}
\caption{Left: the diagram that gives rise to fermionic EDMs via the insertion of the operator $hF\tilde F$ from Eq.~(\ref{Leff}). Right: the two-loop diagram that leads to fermion EDMs in the model involving a VL lepton, $\psi$, coupled to a singlet, $S$, that mixes with the Higgs. The cross on the scalar line indicates that this contribution is proportional to the mixing term, $A$, in the scalar potential.} 
\label{fig:edm_diags}
\end{figure*}

The $CP$-odd Higgs operator (\ref{Leff}) generates fermionic EDMs via a Higgs-photon loop (as seen in Fig.~\ref{fig:edm_diags}),
\begin{align}
 d_i &= \tilde{c}_h \frac{|e|m_f}{4\pi^2\tilde{\La}^2} \ln\left(\frac{\La_{\rm UV}^2}{m_h^2}\right) 
\\
&= d^{(2l)}_{f} \times \frac{\tilde{c}_h}{\alpha/(4\pi)}\times 
\frac{v^2}{\tilde \La^2}\ln\left(\frac{\La_{\rm UV}^2}{m_h^2}\right),
\end{align}
with explicit dependence on the UV scale $\Lambda_{\rm UV}$. If this scale is identified with $ \tilde{\La} $, then 
using the current bound on the electron EDM, $|d_e|< 1.05 \times 10^{-27}e\,$cm \cite{YbF}, we find
\be 
 \tilde{\La} \gtrsim 50 \sqrt{\tilde{c}_h}\,{\rm TeV}.
\ee
Translating this to the Higgs diphoton branching ratio results in  the conclusion that $CP$-odd corrections  are limited by 
\be
 \Delta R_{\gamma\gamma}(\tilde c_h) \lesssim 1.6\times 10^{-4}.
\ee
However, this conclusion can be relaxed in specific UV completions. As we discuss in the next subsection, 
the logarithm $\ln(\tilde{\La}^2/m_h^2)\sim 10$ cannot generally be stretched all the way to 50 TeV, 
as the loops of VL charged particles provide a much lower cutoff, while certain degeneracies may provide
more significant qualitative changes to the implications of EDM limits.
\subsection{UV-complete examples with VL fermions}

\subsubsection{Singlet scalar with pseudoscalar coupling to VL fermions} 

 We will now consider a specific UV completion which allows the full two-loop function to be taken into account for the electron EDM.
The addition of a (hyper)charged VL fermion $\psi$ with mass $m_\psi$ transforming as $(1,1,Q_\psi)$ under SU(3)$\times$SU(2)$\times$U(1), and a singlet $\hat S$ with a Higgs-portal 
interaction with the Higgs doublet $H$ \cite{Hportal}, leads to 
the following Lagrangian:
\begin{align}
  {\cal L} _{SH\ps}&= \bar\ps i\gamma^\mu (i\partial_\mu-  eQ_\psi A_\mu)\ps
\nonumber
\\
&~~~~+ \bar \ps \left[m_\ps + \hat S(Y_S + i  \gamma_5 \tilde{Y}_S)\right]  \ps + {\cal L}_{HS}.
  \end{align}
The terms in ${\cal L}_{HS}$ contain scalar kinetic terms and describe the Higgs-portal interaction between $\hat S$ and $H$ via the following potential:
  \begin{align}
V_{HS} &= -\mu_H^2 H^\dag H + \lambda_H (H^\dag H)^4 + \frac{1}{2} \hat m_S^2 \hat S^2 
\nonumber
\\
&~~~~~~~~+ A H^\dag H \hat S  - B \hat S 
+ \frac{\lambda_S}{4} \hat S^4\,.
\end{align}
$CP$-odd couplings of the Higgs proportional to the combination $A\tilde{Y}_S$ are generated, while the term linear in $\hat S$ can always be adjusted to ensure that $\langle \hat S\rangle = 0$. 
 We retain only the photon contribution of the $J_\mu^\psi$ vector current, as the $Z$ contribution 
is suppressed by the small value of $g_V^e$. After the breaking of $SU(2)\times U(1)$, the $\hat S $ field mixes with what would be 
the SM Higgs boson $\hat h$ to produce two mass eigenstates $h$ and $S$,
\begin{equation}
\left( 
\begin{array}{c}
\hat h \\
\hat S
\end{array} 
\right)
= 
\left( 
\begin{array}{cc}
c_\theta &  s_\theta \\
-s_\theta &  c_\theta
\end{array}
\right)
\left( 
\begin{array}{c}
 h \\
 S
\end{array} 
\right), ~~~~~~~~~ \tan 2\theta = \frac{2 A v }{\hat m_S^2 -2\lambda_H v^2},
\end{equation}
where $s_\theta$ ($c_\theta$) stands for $\sin\theta$ ($\cos\theta$). Both mass eigenstates inherit Higgs-like interactions with the SM fields and couplings to $\psi$ fermions.

The dominant two-loop contribution to fermion EDMs is well-known \cite{BZ}, and specializing to our case we 
arrive at the following result for the electron EDM as a function of $\tilde Y_{S}$, $\theta$, and $m_\psi$:
\be
\label{2loopSh}
 d_f = d_f^{(2l)} \times  Q_\ps^2\tilde{Y}_S \frac{v}{m_\ps} \sin(2\theta)\left[ g(m_\ps^2/m_h^2)-g(m_\ps^2/m_S^2) \right],
\ee
where the loop function is given by
\be 
 g(z) = \frac{z}{2} \int_0^1 dx \frac{1}{x(1-x)-z} \ln\left(\frac{x(1-x)}{z}\right),
 \label{eq:g}
\ee
which satisfies $g(1)\sim1.17$ and $g\sim \frac{1}{2}\ln z$ for large $z$. 
We show the Feynman diagram responsible for this contribution on the right of Fig.~\ref{fig:edm_diags}.

It is instructive to consider different limits of (\ref{2loopSh}). When $m_h\ll m_\psi,m_S$, 
to logarithmic accuracy  $ g(m_\ps^2/m_h^2)-g(m_\ps^2/m_S^2) \to \fr12 \ln(m_{\rm min}^2/m_h^2)$, 
where $m_{\rm min}$ is the smaller of $m_S$ and $m_\psi$. In this limit, the heavy fields can be integrated out sequentially, with 
$S$ and $\psi$ first, and  $h$ second. The first step is simplified by the use of the chiral anomaly equation 
for $\psi$, $  \partial_\mu \bar \psi \gamma_\mu\gamma_5 \psi = 2i\bar \psi \gamma_5 \psi +\frac{\al}{8\pi} Q_\ps^2 F_{\mu\nu}\tilde{F}_{\mu\nu}$. 
This leads to the following identification:
\be
\frac{\tilde c_h}{\tilde \Lambda^2} = \frac{\alpha Q^2_\psi}{4\pi}\frac{\tilde Y_S A}{m_S^2 m_\psi}; ~~ \La_{\rm UV} \simeq {\rm min}(m_S, m_\psi).
\ee
Apart from a smaller value for the logarithmic cutoff, the result in this limit differs little from the contact operator case above. Even if the value of the logarithm 
is not enhanced, $\ln(m_{\rm min}^2/m_h^2)\sim O(1)$, the corrections to the Higgs diphoton rate will be limited to at most the sub-percent level unless 
a fine-tuned cancellation of $d_e$ is arranged with some other $CP$-odd source. 

\begin{figure*}
\includegraphics[scale=0.6]{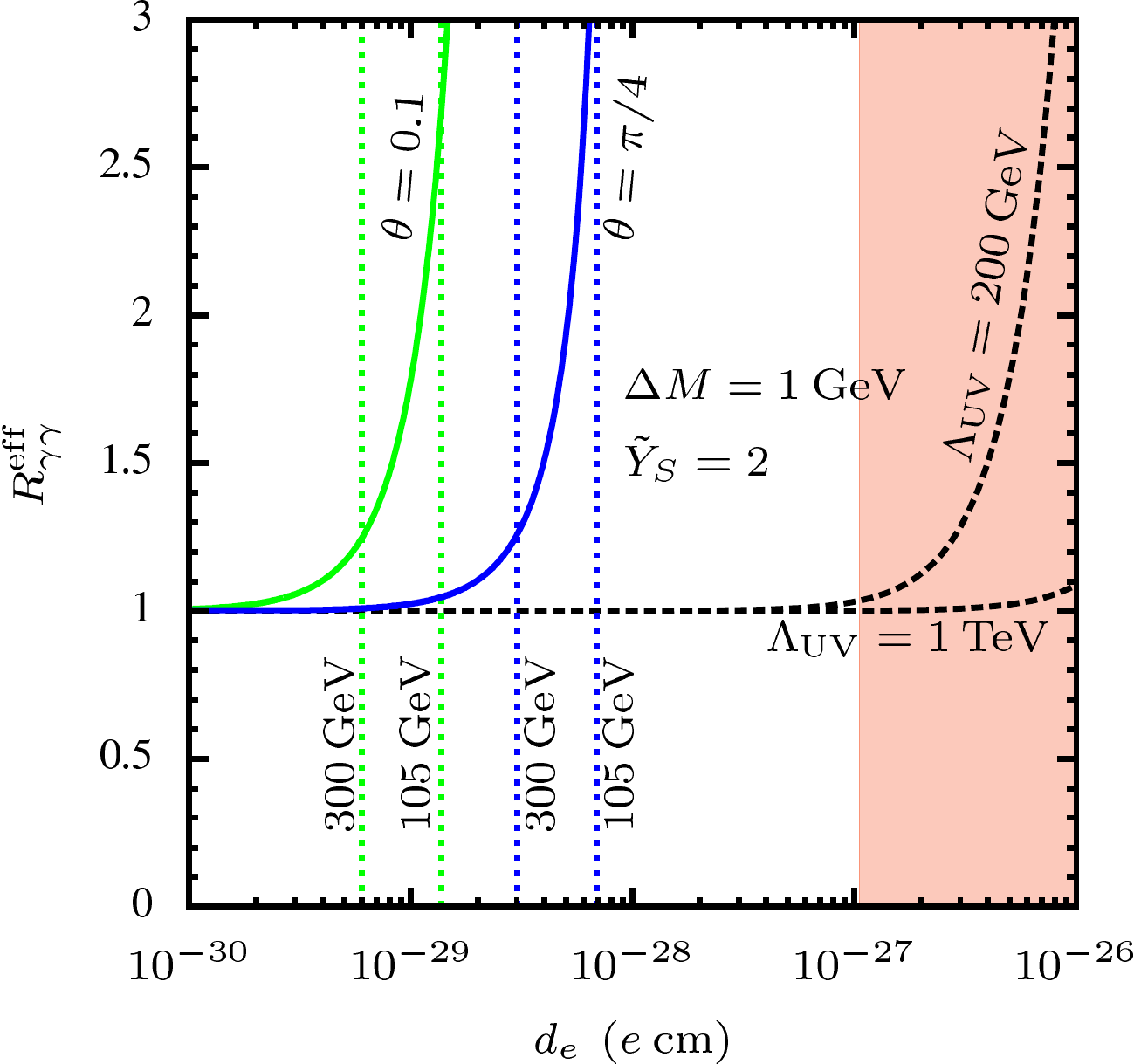}\hspace*{0.5cm}\includegraphics[scale=0.6]{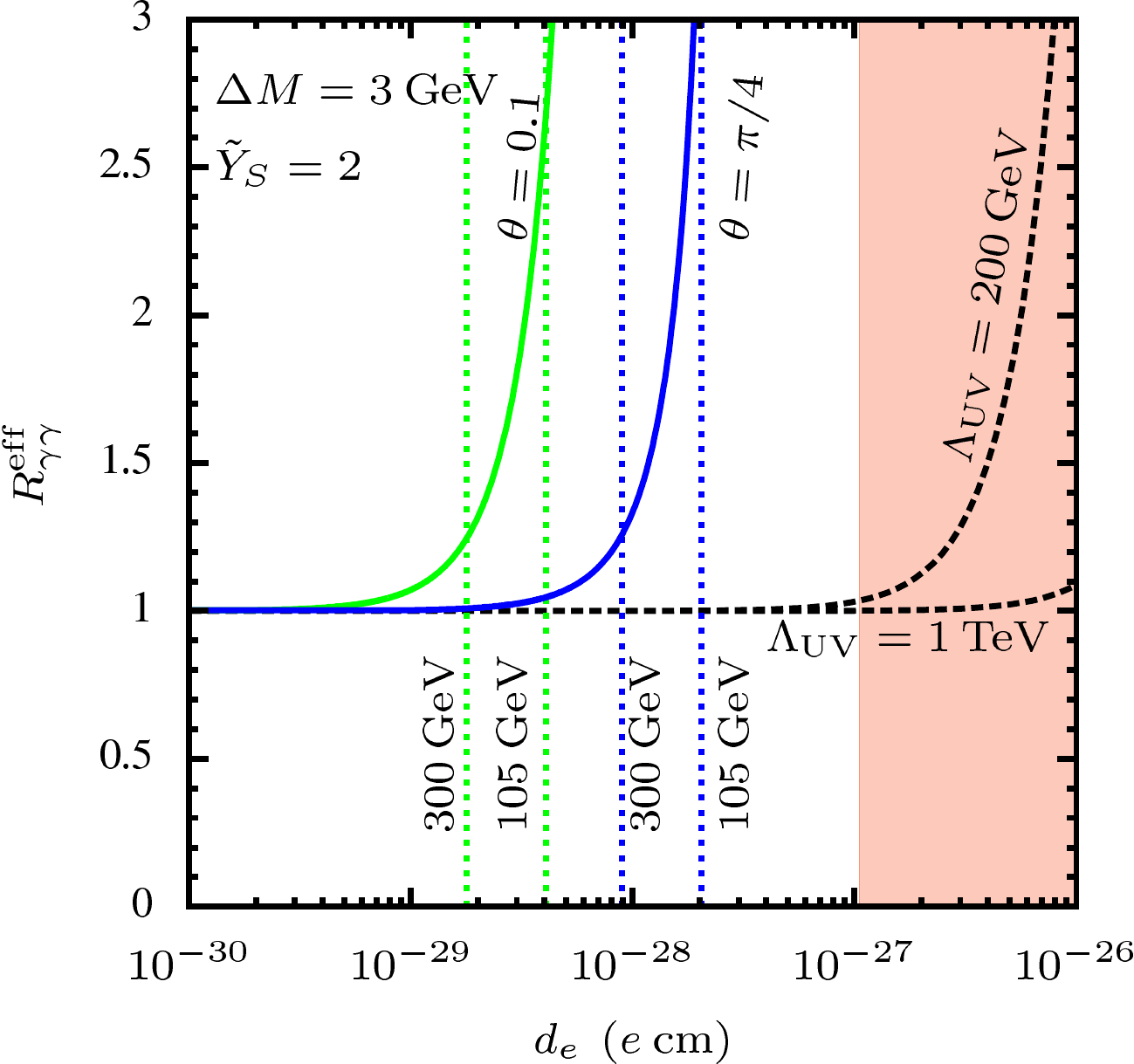}
\caption{The effective increase in the diphoton rate as a function of the electron EDM coming from a coupling of the Higgs to $F_{\mu\nu}\tilde F^{\mu\nu}$.  The black dashed lines show the relationship in the case of the contact operator $hF_{\mu\nu}\tilde F^{\mu\nu}$ simply cut off at the scales $\Lambda_{\rm UV}=200$~GeV and 1~TeV.  The solid lines show the relationship in the case of a scalar singlet, $S$,  nearly degenerate with the Higgs coupled to a VL fermion, $\psi$.  We choose a splitting between $m_S$ and $m_h$ of $\Delta M=1$~GeV (left panel) and 3~GeV (right panel) and a $CP$-odd Yukawa coupling of the singlet to the VL fermions of $\tilde Y_S=2$.  The curve on the left of each panel (green) is for a mixing angle $\theta=0.1$ and that on the right of each panel (blue) for $\theta=\pi/4$.  The dotted lines show the value of $d_e$ implied for the two mixing angles for $m_\psi=105$~GeV and 300~GeV.  Values of the electron EDM that are excluded experimentally, $d_e>1.05\times10^{-27}~e$~cm,  are in the shaded region.  We observe that the degenerate scalar allows for a sizable apparent increase in the Higgs diphoton rate in the $CP$-odd channel while not conflicting with the electron EDM limit, unlike the simple contact operator case.} 
\label{fig:R_vs_edm}
\end{figure*}

We now consider a different near-degenerate limit, $|m_h - m_S| \ll m_{h}$, which turns out to be more interesting as it allows 
the EDM constraints to be bypassed. 
If the difference between the masses is small, we can approximate
\be
\sin(2\theta)(m_S^2-m_h^2) \to 2Av, 
\ee
and the EDM becomes
\begin{align}
\label{limit2}
d_f &= d_f^{(2l)} \times  Q_\ps^2\tilde{Y}_S \fr{2Av^2m_\psi}{m_h^4}  g'(m_\psi^2/m_h^2) 
\\
&\longrightarrow~ d_f^{(2l)} \times 
 Q_\ps^2\tilde{Y}_S   \fr{Av^2}{m_h^2m_\psi} ,
\end{align}
where in the final step we made use of  the large $m_\psi$ limit. 

The limiting case (\ref{limit2}) receives no logarithmic enhancement. Moreover, the value of the $A$ parameter can be very small, 
comparable to the mass splitting between $h$ and $S$ or less. An $O({\rm 1~GeV})$ mass splitting would naturally place 
$ Av^2/(m_h^2m_\psi)$ in the $O(10^{-2}-10^{-3})$ range, suppressing the EDM safely below the bound.

At the same time, as explicitly shown in Ref.~\cite{bmp}, modifications to the $h\to \gamma \gamma$  rate can be significant, 
and  enhancement can come from the $ F_{\mu\nu}\tilde F^{\mu\nu}$  amplitude.  Unlike corrections to the
$F_{\mu\nu}  F^{\mu\nu}$ amplitudes that can enhance or suppress the effective rate,  the $CP$-odd channel always 
adds to $R_{\gamma\gamma}$.  Assuming that the mass difference between the singlet and the Higgs is small enough that 
they cannot be separately resolved (which requires $\left|m_S-m_h\right|\lsim 3$~GeV with current statistics~\cite{bmp}), the apparent increase in the diphoton rate in this model is
\begin{align}
\label{Rvalue}
R_{\gamma\gamma}^{\rm eff} (\tilde{Y}_S)&= \cos^2\theta \times \frac{ {\rm Br} _{h\rightarrow \gamma\gamma}}{ {\rm Br} _{h\rightarrow \gamma\gamma}^{\rm SM}}+ \sin^2\theta \times \frac{ {\rm Br} _{S\rightarrow \gamma\gamma}}{ {\rm Br} _{h\rightarrow\gamma\gamma}^{\rm SM}}.
\end{align}
If $\theta$ is in the range
\begin{align}
\sqrt{\frac{\Gamma_{\hat S\to\gamma\gamma}}{\Gamma_{\hat h\to\gamma\gamma}}{\rm Br}_{h\to\gamma\gamma}^{\rm SM}}\lsim\theta\lsim\sqrt{\frac{\Gamma_{\hat h\to\gamma\gamma}}{\Gamma_{\hat S\to\gamma\gamma}}}
\end{align}
and $\Gamma_{\hat h\to\gamma\gamma}\sim\Gamma_{\hat S\to\gamma\gamma}$ then $R_{\gamma\gamma}$ simplifies to a $\theta$-independent expression,
\begin{align}
R_{\gamma\gamma}^{\rm eff}(\tilde{Y}_S)&\simeq 1+\frac{\Gamma_{\hat S\to\gamma\gamma}}{\Gamma_{\hat h\to\gamma\gamma}}.
\end{align}
The rate for the weak eigenstate $\hat S$ to decay to two photons via its pseudoscalar coupling to the VL fermions is
\begin{align}
\Gamma_{\hat S\to\gamma\gamma}=\frac{\alpha^2 Q_\psi^4 \tilde Y_s^2 m_S^3}{256\pi^3m_\psi^2}\left|A_{1/2}^P\left(\frac{m_S^2}{4m_\psi}\right)\right|^2,
\end{align}
with
\begin{align}
A_{1/2}^P\left(\tau\right)=\frac{2}{\tau}\left(\sin^{-1}\sqrt{\tau}\right)^2.
\end{align}
For large $m_\psi$ the apparent diphoton increase can then be expressed as
\begin{align}
R_{\gamma \gamma}^{\rm eff}(\tilde{Y}_S)\sim1+
Q_\psi^4\left(\frac{\tilde Y_S}{2}\right)^2\left(\frac{150~{\rm GeV}}{m_\psi}\right)^2.
\end{align}
A sizable increase in the apparent diphoton rate is seen to require rather large Yukawa couplings or light VL fermions.  The VL leptons must be heavier than 105~GeV to avoid limits from LEP.  Their decay channels are fairly model-dependent but they are well within the reach of the LHC if they are at all relevant for the $h\to\gamma\gamma$ rate.  For more discussion on experimental searches for such VL fermions, see~\cite{bmp}.

In Fig.~\ref{fig:R_vs_edm} we show the relationship between the electron EDM and the enhancement to the Higgs diphoton rate that comes from the operator $hF_{\mu\nu}\tilde F^{\mu\nu}$ for both the contact operator and nearly degenerate singlet cases.  In the case of the contact operator, we show two cutoffs, $\Lambda_{\rm UV}=200$~GeV and 1~TeV.  As seen in Sec.~\ref{sec:contact_op_lim}, it is apparent that in this simple situation, any appreciable increase in the $h\to\gamma\gamma$ rate must be accompanied by a value of the electron EDM that is in conflict with the present experimental limit.  We also show the relationship between $R_{\gamma\gamma}^{\rm eff}$ and $d_e$ in the singlet case for two values of the mixing angle, $\theta=0.1$ and $\pi/4$, fixing the pseudoscalar Yukawa to $\tilde Y_S=2$ and choosing $Q_\psi=1$.  Different values of $R_{\gamma\gamma}^{\rm eff}$ and $d_e$ then correspond to different values of $m_\psi$.  It is now apparent that a sizable increase in the effective diphoton rate can be obtained in this model without inducing a value of the electron EDM that is presently excluded, demonstrating a UV completion of the effective interaction that evades the constraints implied by a simple analysis of this contact operator.  The reason that the EDM constraints are evaded in this case is clear:  mixing of the 
two fields, $\hat h$ and $\hat S$, due to the small mass difference can proceed rather efficiently even with a small value of $A$, while the EDM loop diagrams do not enjoy the same resonant enhancement.
In this model, for fixed $R_{\gamma\gamma}^{\rm eff}$, $d_e$ increases with increasing $\Delta M$ and $\sin 2\theta$. The rough upper limit on $\Delta M$ of around 3~GeV with current data implies an upper limit on $d_e$ of $\sim 10^{-28}~e~$~cm for $R_{\gamma\gamma}^{\rm eff}\simeq1.5-2$. Separately resolving a degeneracy near 125~GeV or limiting the size of a potential mass splitting with more data clearly has important implications for EDM searches.

\subsubsection{Full VL generation with $CP$-violating Higgs couplings.} 

Another simple UV completion is a full VL generation of SM-like fields $E_R\sim(1,1,-1)$ and $L_L\sim(1,2,-1/2)$, 
with their mirror image fields $E_L$ and $L_R$,
\be
\label{EL}
-{\cal L}_{EL} \supset  (\bar E_L, ~ \bar L_L)
 \left(
\begin{array}{cc}  M_E & y_1 H \\y_2 H^* &M_L
\end{array}
 \right)
 \left( \begin{array}{c}E_R \\L_R\end{array} \right)+{\rm h.c.}
\ee

Every entry in this mass matrix, $M_{E(L)}, y_{1(2)}$, can be complex. However, there is only one physical $CP$-odd phase combination 
that cannot be removed by a field redefinition $\sim \phi_E+\phi_L- \phi_1-\phi_2$, which will appear in Higgs-fermion $CP$-odd vertices. 
\begin{widetext}
For the purposes of calculation, it is more convenient to switch to the mass eigenstate basis for the $Q=1$ fermions (we denote
the masses $m_1$ and $m_2$),  related to the original basis (\ref{EL}) by a unitary rotation of the left- and right-handed fields:
\be
\left(
\begin{array}{cc}  M_E & y_1 v/\sqrt{2}\\y_2 v/\sqrt{2} &M_L
\end{array}
 \right)
=\left(
\begin{array}{cc}  \cos\theta_L & \sin\theta_L e^{i\phi_L}\\-\sin\theta_Le^{-i\phi_L} &\cos\theta_L
\end{array}
 \right)
\left(
\begin{array}{cc}  m_1 & 0 \\0 &m_2
\end{array}
 \right)
\left(
\begin{array}{cc}  \cos\theta_R & -\sin\theta_R e^{-i\phi_R}\\\sin\theta_Re^{i\phi_R} &\cos\theta_R
\end{array}
 \right).
\ee
In the mass eigenstate basis, the Higgs fields develops the following couplings to the $\psi_1$ and $\psi_{2}$ fermions:
\ba
\label{h12} 
{\cal L} &= &\fr{h}{2v} m_1\bar\psi_{1L}\left[1 - \cos(2 \theta_L) \cos(2 \theta_R) - 
 \frac{m_2}{m_1}e^{-i (\ph_L - \ph_R)} \sin(2 \theta_L) \sin(2 \theta_R) \right]\psi_{1R} \\
&+&\fr{h}{2v} m_2\bar\psi_{2L}\left[ 1 - \cos(2 \theta_L) \cos(2 \theta_R) - 
 \frac{m_1}{m_2}e^{i (\ph_L - \ph_R)}  \sin(2 \theta_L) \sin(2 \theta_R)\right] \psi_{2R} +{\rm h.c.}+\cdots
\nonumber
\ea
\end{widetext}
The ellipsis denotes the off-diagonal $h\bar \psi_1\psi_2$ couplings, which will not affect the EDMs or Higgs decay phenomenology
within our approximations. 
The $CP$-odd vertices from this Lagrangian can now be inserted directly into the two-loop formulae,
\begin{align}
d_e^{h\gamma} &= d_e^{(2l)} \times  
 \sin(\ph_L - \ph_R)\sin(2 \theta_L) \sin(2 \theta_R)\nonumber\\
  & \;\;\;\;\;\;\;\;\;\;\; \times \fr{m_1m_2}{m_h^2} \left[ \frac{g(z_1)}{z_1}-\frac{g(z_2)}{z_2}\right],
\end{align}
where $z_i = m^2_{i}/m_h^2$.
In addition to the $h\gamma $ two-loop diagram, there is also a $WW$ two-loop contribution, with the same topology. 
The mass of the neutral fermion that enters this diagram is given by $M_L$, where
\begin{align}
|M_L|^2 &= m_2^2 \cos^2\theta_L \cos^2\theta_R + m_1^2 \sin^2\theta_L \sin^2\theta_R \nonumber\\
 & \;\;\;\; + 
 \frac{m_1 m_2}{2} \cos(\phi_L - \phi_R) \sin(2\theta_L)\sin(2 \theta_R).
\end{align}
$CP$ violation enters the $WW$ diagram via the relative phase of the left- and right-handed charged currents. Performing the calculation, we find 
\begin{align}
d_e^{WW} &= d_e^{(2l)} \times  
 \sin(\ph_L - \ph_R)\sin(2 \theta_L) \sin(2 \theta_R) \nonumber\\
  & \;\;\; \times \fr{m_1m_2}{m_W^2} \fr{\alpha_W}{8\alpha}
\left[ \frac{j(z_1,z_L)}{z_1}-\frac{j(z_2,z_L)}{z_2}\right],
\end{align}
where $z_i = m^2_{i}/m_W^2$, $z_L = |M_L|^2/m_W^2$, $\alpha_W = g_W^2/(4\pi)$, and the new loop function 
$j$ is defined as 
\be
j(z,r)=z\int_0^1\frac{dx(1-x)}{(x-z)(1-x)-rx} \ln\left(\frac{x(1-x)}{z(1-x)+rx}\right).
 \label{eq:j}
\ee
Calculations of these two-loop effects closely resemble those for the  
two-loop chargino-neutralino EDM contributions in ``split SUSY" models \cite{AGR}
and the two-loop EDMs in theories with additional $CP$-violation in the 
top-Higgs coupling (see, e.g.,~\cite{HPR}). 

\begin{widetext}
In this model, the increase in the Higgs diphoton decay rate resulting from $CP$-violating couplings is
\begin{align}
R_{\gamma\gamma}(\ph_L - \ph_R)&= 1+\left(\frac{d_e}{d_e^{\left(2l\right)}}\right)^2\frac{\left|m_2^2A^P_{1/2}\left(m_h^2/4m_1^2\right)-m_1^2A^P_{1/2}\left(m_h^2/4m_2^2\right)\right|^2}{4m_1 m_2\left|A_{\rm SM}\right|^2D},
\end{align}
\end{widetext}
where $D$ is a (typically ${\cal O}(1)$) combination of two-loop functions,
\begin{align}
D&=\fr{m_1m_2}{m_h^2} \left[ \frac{g(z^h_1)}{z^h_1}-\frac{g(z^h_2)}{z^h_2}\right] \nonumber\\
 & \;\;\; +\fr{m_1m_2}{m_W^2} \fr{\alpha_W}{8\alpha}
\left[ \frac{j(z^W_1,z^W_L)}{z^W_1}-\frac{j(z^W_2,z^W_L)}{z^W_2}\right].
\end{align}
A large enhancement of the diphoton rate through $CP$-violating effects would require large mass splittings between\footnote{We note that a large splitting is problematic for electroweak precision measurements but a detailed analysis of this issue lies outside the scope of this paper.  Studies of electroweak precision and an increase in the Higgs diphoton rate have recently been undertaken in, e.g.,~\cite{VL_EWP,bmp}.} $\psi_1$ and $\psi_2$ and for $d_e/d_e^{\left(2l\right)}$ to be at least a factor of a few.  Since $d_e^{\left(2l\right)}$ is itself larger than the present limit on the electron EDM, a sizable $CP$-odd enhancement to the $h\to\gamma\gamma$ rate in this model will generate an electron EDM in conflict with experiment.  Therefore, this model is an example of a UV completion that gives rise to the operator $hF_{\mu\nu}\tilde F^{\mu\nu}$ whose behavior aligns with that of the simple contact operator in Sec.~\ref{sec:contact_op_lim}: a large $CP$-odd contribution to the Higgs diphoton rate conflicts with the experimental limit on the electron EDM.

\section{CP and flavor observables vs dilepton decays}
\label{sec:flavor}
We now turn our attention to the leptonic branching ratio of the Higgs, and the interplay with sources of flavor violation
in the Higgs couplings \cite{Hflavor,Hflav_rev}. We will 
consider specific tree-level dimension-six threshold corrections to the lepton Yukawa couplings \cite{HPR,Hcube,lebedev}, assumed to arise, e.g.,
from a lepton-specific extension to the Higgs sector,
\begin{align}
 -{\cal L} &= Y^{ij} \bar{L}_L^i H e_R^j + \frac{Z^{ij}}{\La^2} (H^\dagger H) \bar{L}_L^i H e_R^j+{\rm h.c.}
 \label{eq:L_lepton}
   \\
  &\rightarrow m_i \bar{l}_i l_i +\frac{m_i}{v} \bar{l}_i h (\de_{ij} + \al_{ij} + i\beta_{ij} \gamma^5) l_j +\cdots
\end{align}
where the second line refers to the mass eigenstate basis, and the normalization assumes $m_i > m_j$. The flavor matrix
is
\begin{align}
\al_{ij} +i\beta_{ij} = \frac{v^3}{\sqrt{2}\Lambda^2m_i}\left(U Z V^\dagger\right)_{ij}
\end{align}
where $U$ and $V$ rotate the left- and right-handed lepton fields from the weak basis to the mass basis, respectively.

\begin{table*}
\begin{tabular}{|c|c|c|} \hline
observable & source & limit \\ \hline
Br$(\mu \rightarrow e\gamma) <  2.4\times 10^{-12}$ & MEG \cite{MEG} & $\gamma_{12} <4.7\times10^{-3}$ \\
Br$(\tau \rightarrow e \gamma) <  3.3\times 10^{-8}$& BaBar \cite{Babar} & $\gamma_{13} <1.4$ \\
Br$(\tau \rightarrow \mu \gamma) <  4.4\times 10^{-8}$& BaBar \cite{Babar} & $\gamma_{23} < 1.6$ \\
$\Gamma(\mu N \rightarrow eN)/\Gamma_{\rm capture} < 7\times 10^{-13}$ & SINDRUM-II \cite{Sindrum} & $\gamma_{12} < 6\times10^{-2}$ \\
Br$(\tau \rightarrow \mu K^+K^-) <  6.8\times 10^{-8} $ & Belle \cite{BelleKK} &  $\gamma_{23} < 57$ \\
Br$(\tau \rightarrow 3\mu) < 2.1\times 10^{-8}$ & Belle \cite{Belle3mu} & $\gamma_{23} < 67$ \\ \hline
\end{tabular}
\caption{Limits on the flavor matrix elements $\gamma_{ij}\equiv\sqrt{\al_{ij}^2+\beta_{ij}^2}$ from various observables.}
\label{tab:LFV}
\end{table*}

The correction to the Higgs dilepton branching ratio takes the form,
\be
B_l = \frac{\Gamma_{l\bar{l}}}{\Ga_{l\bar{l}}^{\rm SM}} \simeq |1 + \al_{ll}|^2 + |\beta_{ll}|^2,
\ee
so that any $CP$-odd correction is again necessarily positive.

\subsection{Flavor sensitivity}

We assume a generic flavor structure for the matrices $\al_{ij}$ and $\beta_{ij}$. Integrating out the Higgs, the operators with
minimal Yukawa suppression are two-loop transition dipoles with top and $W$ loops~\cite{Hflavor},
\be
 {\cal L}_{\rm dipole}=\frac{\al e}{32\pi^3} \frac{m_i}{v^2} \left(C_t+C_W\right)\bar{l}_i F\sigma\left(\al_{ij} +i\beta_{ij}\gamma^5\right)  l_j.
\ee
The loop functions are\footnote{We include only the leading diagrams involving virtual $W$/Goldstone bosons.}
\begin{align}
C_t&=2 N_c Q_t^2f\left(z_t\right),
\\
C_W&=-\left\{3f\left(z_W\right)+\frac{23}{4}g\left(z_W\right)\right.
\nonumber
\\
&~~~~~~~~\left.+\frac{3}{4}h\left(z_W\right)+\frac{1}{2z_W}\left[f\left(z_W\right)-g\left(z_W\right)\right]\right\},
\end{align}
where $z_t=m_t^2/m_h^2$, $z_W=m_W^2/m_h^2$, $g\left(z\right)$ is defined in Eq.~(\ref{eq:g}), and
\begin{align}
f(z) &= \frac{z}{2} \int_0^1 dx \frac{1-2x(1-x)}{x(1-x)-z} \ln\left(\frac{x(1-x)}{z}\right),
 \\
h(z) &= z^2\frac{\partial}{\partial z}\left(\frac{g\left(z\right)}{z}\right).
\end{align}
There are also one-loop contributions to the dipoles with an $h-\ta$ loop, proportional (with our normalization) to $Y_\ta^2$, and Higgs-mediated four-fermion interactions that are further Yukawa-suppressed.

Using the effective interactions arising from~(\ref{eq:L_lepton}), various LFV transition rates are straightforwardly computed as discussed e.g. in \cite{brmmr,lebedev},
and we summarize some of the stronger limits in Table~\ref{tab:LFV}. The transition dipoles generally lead to the strongest constraints, despite being
generated at two-loop order, as they are not subject to additional Yukawa suppression. Indeed,  the largest contribution
to $\mu\rightarrow e$ conversion actually comes from the induced $\mu e\gamma$ vertex, despite being loop-suppressed
relative to the Higgs-mediated four-fermion operator. 

We observe that most of the LFV limits are relatively weak, particularly in the $\ta$ sector, and so thresholds that
impact BR$(h\rightarrow \ta\ta)$ could still introduce new flavor structures in the $\ta$ sector. The most stringent
limits apply to $\al_{12}$ and $\beta_{12}$, and a generic flavor structure in the muon sector would limit branching ratio corrections
to the percent level.

\subsection{CP sensitivity}

There are analogous two-loop contributions to the electron EDM,
\begin{align}
 d_e = \frac{\al e}{16\pi^3} \frac{m_e \beta_{11}}{v^2} \left(C_t+C_W\right).
\end{align}
The current electron EDM bound~\cite{YbF} implies $\left|\beta_{11}\right|<0.13$. Taken as a generic flavor-independent limit, this
 would restrict any CP-odd corrections to the branching ratio to ${\cal O}(2\%)$.
 
 \subsection{Comments on UV completions}

A simple candidate model that can give rise to the effective Higgs-lepton 
interactions in (\ref{eq:L_lepton}) is a two Higgs doublet model~\cite{Branco:2011iw} with one 
doublet, $H_q$, coupled to quarks and the other, $H_\ell$, to leptons. 
$H_q$ can be arranged to be SM-like, suppressing the branching of the SM-like 
Higgs to leptons.  The charged Higgses, if light, could contribute to the 
branching rate to diphotons.  However, substantially increasing this rate 
appears to require large, negative quartic couplings, with potential issues 
for vacuum stability. For further discussion of this point in the context of
colored particles contributing to the diphoton rate, see~\cite{Reece:2012gi}.

\bigskip

The relatively weak limits on flavor-violating observables could allow for 
${\cal O}(1)$ deviations in the Higgs sector with respect to leptons. 
Measuring the Higgs decay rate to taus would be a highly desirable step
towards testing this possibility. Looking for lepton flavor-violating decays with a large sample of taus, as could be obtained at a Super-$B$ factory, would shed further light on the situation.  Additionally, we note that it appears possible to 
check the $CP$ properties of the $h-\tau-\tau$ coupling at a linear 
collider~\cite{Berge:2012wm}. 

\section{Concluding Remarks}
\label{sec:conclusions}
 With the discovery of a new Higgs-like resonance at the LHC, attention is turning to precision tests
 of its interactions. The 
 variety of decay channels
 accessible at $\sim125$~GeV is already providing important information about its couplings to vector
 bosons and fermions. Further tests of these production and decay channels in coming years will provide an important new probe of
 physics beyond the SM,  and allow for a useful interplay with other precision data, particularly in the Yukawa
 sector. In this paper, we have  studied the extent to which a generic new threshold with $CP$ and lepton flavor violation
 can impact Higgs branchings to diphotons and dileptons. We find that precision constraints on EDMs
 and LFV decays restrict this possibility quite significantly in many cases. In particular, large $CP$-violating
 contributions to $h\rightarrow \gamma\gamma$ require an extended scalar sector with mass degeneracies. While
 there is currently limited information about $h\rightarrow \ta\ta$ decays, large corrections to the SM rate
 are possible with new flavor structures at relatively low scales. Progress in studies of rare $\ta$ decays,
 e.g., at a Super-$B$ factory, could provide further constraints on this possibility.
 
 {\it Note:} As this work was being finalized, several related publications appeared on the arXiv. Reference~\cite{MV} discusses $CP$-odd
 contributions to Higgs digamma decays in models with VL fermions, while Refs.~\cite{new_param} consider 
 the $CP$ properties of the Higgs-like resonance.
 We also thank P.~Winslow for informing us of related work in progress with S.~Tulin. 

\begin{acknowledgements}
We would like to thank B.~Batell for helpful discussions.  The work of  D.M., M.P. and A.R. is supported 
in part by NSERC, Canada, and research at the Perimeter Institute is supported in part by the Government 
of Canada through NSERC and by the Province of Ontario through MEDT. 
\end{acknowledgements}

\end{document}